# Observation of Coulomb gap in the quantum spin Hall candidate single-layer 1*T'*-WTe$_2$

Ye-Heng Song[1†], Zhen-Yu Jia[1†], Dongqin Zhang[1], Xin-Yang Zhu[1], Zhi-Qiang Shi[1], Huaiqiang Wang[1], Li Zhu[1], Qian-Qian Yuan[1], Haijun Zhang[1,2], Dingyu Xing[1,2], Shao-Chun Li[1,2*]

1. National Laboratory of Solid State Microstructures, School of Physics, Nanjing University, Nanjing 210093, China

2. Collaborative Innovation Center of Advanced Microstructures, Nanjing University, Nanjing 210093, China

† These authors contributed equally to this work.

* email: scli@nju.edu.cn

The two-dimensional topological insulators (2DTI) host a full gap in the bulk band, induced by spin-orbit coupling (SOC) effect, together with the topologically protected gapless edge states. However, the SOC-induced gap is usually small, and it is challenging to suppress the bulk conductance and thus to realize the quantum spin Hall (QSH) effect. In this study, we find a novel mechanism to effectively suppress the bulk conductance. By using the quasiparticle interference (QPI) technique with scanning tunneling spectroscopy (STS), we demonstrate that the QSH candidate single-layer 1T'-WTe$_2$ has a semimetal bulk band structure with no full SOC-induced gap. Surprisingly, in this two-dimensional system, we find the electron interactions open a Coulomb gap which is always pinned at the Fermi energy ($E_F$). The opening of the Coulomb gap can efficiently diminish the bulk state at the $E_F$ and is in favor of the observation of the quantized conduction of topological edge states.

The 2DTIs show great potentials in future applications, such as low dissipation electronics and quantum computing[1-4]. Since the discovery of quantum spin Hall (QSH) effect in HgTe/CdTe quantum wells [5,6], enormous efforts have been devoted to practically useful 2DTI materials [7-17]. However, most of the discovered 2DTI materials so far have only a small SOC-induced gap, which leads to serious difficulties for experimental realization of their intrinsic QSH effect. One big challenge is to delicately tune the Fermi level within the energy gap and to separate the edge conductance channels from the bulk. It is even more difficult for the QSH candidates with semimetal band structure, which host nontrivial direct gaps in the whole Brillion zone, but are lack of a full gap. Therefore, how to efficiently diminish the bulk conductance is of great significance to to achieve the quantized conduction measurement of the edge conductance channels.

Qian *et al* [9] predicted a class of QSH materials in the single-layer $1T$'-phase of transition metal dichalcogenide (TMD), $TX_2$, where T represents a transition metal atom (Mo, W) and X stands for a chalcogen atom (S, Se or Te). The band inversion happens between transition metal $d$ orbitals and chalcogenide $p$ orbitals, and the SOC interaction further opens a fundamental band gap [9]. Recently, great experimental progress have been made in the single-layer $1T$'-WTe$_2$ [18-20]. For example, the transport measurement on the exfoliated monolayer WTe$_2$ sheet showed the existence of topological edge conductance [18]. The natural single-layer $1T$'-WTe$_2$ was successfully grown by molecular beam epitaxy (MBE) technique [19,20], and its topological edge states were also visualized by STM/STS measurement[19,20].

However, the experiments showed that the single-layer $1T$'-WTe$_2$ exhibits an insulating behavior at low temperature [18-20], inconsistent with the semimetal bulk band structure as initially predicted by DFT under single-electron frame [9]. Even though several possible mechanisms have been suggested to explain this insulating behavior[19-22], the issue still remains controversial because of the lack of fully understanding of its electronic structure. The STS-QPI technique is suitable and crucial to solve such a controversy, since it has the capability to characterize the band structure with high energy-resolution, for both the occupied and unoccupied states near the Fermi energy $E_F$.

In this study, we employ the QPI-STS/STM to detailedly investigate the electronic structure of single-layer 1T'-WTe$_2$. At first, we clarify that the conduction bands (Te $p$) cross the Fermi level along the Y-Γ-Y direction, and the energy of their bottoms is lower than the top of the valence band (W $d$), in agreement with the DFT calculation under the single-electron frame [9]. Second, we explicitly reveal that there is a Coulomb gap at the Fermi level, which arises from the electron interactions in the 2D system rather than the spin-orbit coupling (SOC). Unlike the SOC-induced gap, as generally considered in 2DTIs, the Coulomb gap discovered in this study always locates at the Fermi level, independent of the electron doping. This exotic gap in the single-layer 1T'-WTe$_2$ can efficiently filter its topological edge channels directly from the vanishing bulk states at the Fermi level, regardless of the gap size.

The single-layer 1T'-WTe$_2$ possesses a sandwich structure with three atomic layers of Te-W-Te [23,24], as shown in Fig. 1a. The 1T'-WTe$_2$ phase is formed due to the spontaneous lattice distortion in the 1T structure where the W-Te-W stacks in the rhombohedral (ABC) atomic-layer order. The distorted W atoms in the $x$ axis form the one-dimensional (1D) zig-zag atomic chains along the $y$ axis and a doubling 2×1 periodicity [25]. It should be noted that the topmost Te atoms are not in a plane due to the distortion of the W atoms underneath. The atomic resolution STM image of the single-layer 1T'-WTe$_2$, as shown in Fig. 1b, discloses the exact atomic registry and the apparent atomic height, which is consistent with the 1T' phase atomic structure. The corresponding reciprocal Brillouin Zone is shown in Fig. 1c.

The local density of states (LDOS) in a wide bias range (U = ±1.0 V), as represented by the differential conductance dI/dV spectrum in supplementary materials Fig. S1, agrees well with the previous reports[19,20] and the DFT calculated band structure for the freestanding monolayer [9]. A series of differential conductance dI/dV spectra (128 curves in total) in a smaller bias range, taken along a line of ~16 nm, are plotted together in Fig. 1d. The key features in these dI/dV curves are the energy gap (red arrow) at the $E_F$ and the kink at ~ -60 mV (blue arrow). These two features are uniform and unchanged in real space. As will be discussed later, the energy gap is always located at the $E_F$ and caused by the Coulomb interaction between electrons. The features that vary along the surface come from the QPI, see particularly the bumps in the positive bias region in Fig. 1d. The dI/dV maps are measured over the 1T'-WTe$_2$ terrace, and two typical maps are depicted in Figs. 1e and 1f, from which the spatial modulations due to the QPI of electronic Bloch waves can be clearly identified.

Figures 2a-2d show the fast Fourier transform (FFT) of the dI/dV maps taken at different bias voltages, sweeping from the unoccupied to the occupied states. More QPI patterns can be found in the supplementary materials Fig. S2. The wave vector $q = k_f - k_i$ obtained from the QPI pattern can be understood as the elastic scattering of electronic Bloch waves, from the initial state of $k_i$ to the final state of $k_f$. In the First Brillion Zone of Figs. 2a and 2b (the middle part), there exist three ellipses in all the QPI patterns, with one located at the center and the other two symmetrically located along the Y-Γ-Y direction, as guided by the pink and green ovals. In Figs. 2c and 2d, two extra features start to appear and are symmetrically located along the Y-Γ-Y as well, as marked by the orange oval. The replicas of these features with weaker intensity are found in the second Brillion Zone, i.e. the upper and lower parts in Figs. 2a-2d. Figure 2e shows the schematic band structure derived from the previous DFT calculation (see Fig. S1F of Ref. [9]). Comparing Figs. 2a-2d with Fig. 2e, one can find that the central pocket comes from two scattering channels: the intra-band scattering of the conduction band (pink oval in Fig. 2a, $q_1$ in Fig. 2e) and that of the valence band (red oval in Fig. 2d, $q_4$ in Fig. 2e). These two channels are distinguishable when the energy is far away from the $E_F$, and entangled with each other when close to each other. The size of the two side ellipse pockets (green ovals) shrinks with decreasing bias voltage, confirming that these ellipses are associated with the electron-like pocket scattering. According to Fig. 2e, they can be assigned as the scattering channel of $q_2$, *i.e.*, the inter-conduction band scattering. The extra features as marked by the orange oval can be assigned as the inter-band scattering between the valence and conduction bands ($q_3$ in Fig. 2e).

The E-q dispersion, namely the scattering band structure, can be extracted from the line cuts taken on the QPI patterns along some specific directions. In Fig. 2f is plotted the E-q dispersion along the Y-Γ-Y direction. The QPI data that are used to extract the E-q dispersion are shown in the supplementary materials, Figs. S2 and S3. For guiding eyes, the black lines are drawn in Fig. 2f to track the dispersions of q2, q3 and q4, respectively. One can see that there exists an energy overlap region between the valence band and the conduction bands, which suggests a semimetal band structure.

To further understand the physical origins for the QPI wave vectors, we compare the experimental results with our DFT simulated patterns. Two typical zoom-in QPI patterns are shown in Figs. 3a and 3b, with the former energy cutting only the conduction band and the latter one cutting the band overlap region. The corresponding schematic constant energy contours (CECs) are plotted in Figs. 3c and 3d, from which the simulated QPI patterns are obtained in terms of the DFT calculation, as shown in Figs. 3e and 3f. In both cases, the experimental QPI results are perfectly consistent with the simulated QPI patterns. Such a good agreement further confirms the validity of the band structure model of Fig. 2e and the origins of these QPI features. In particular, the existence of $q_3$ corresponding to the scattering between the valence and conduction bands provides a decisive evidence for the semi-metallic band structure without a full band gap. This semimetal band structure of single-layer 1T'-WTe$_2$ is also in good agreement with the DFT calculation under single-electron frame [9]. As a result, such experimental results unambiguously show that the insulating behavior in single-layer 1T'-WTe$_2$ is not caused by a single-particle band gap.

We next turn to investigate the puzzling gap at the Fermi level. Our tunneling spectroscopy study confirms that there is a soft gap at the Fermi level. However, the above QPI analysis indicates that the band structure of single-layer 1T'-WTe$_2$ is semimetal, without a full gap between the conduction and valance bands. We thus believe that such a gap at the Fermi level is not attributed to the SOC-induced single-particle gap. The possibility of the superconducting gap, should be excluded, because the previous transport study indicated an insulating behavior at low temperature [20]. Furthermore, the gap cannot be suppressed by applying magnetic field (supplementary materials Fig. S4). To clarify the mechanism for the gap formation, we purposely dope electrons to the single-layer 1T'-WTe$_2$ surface via potassium (K) deposition, to tune the position of Fermi level. Figures 4a and 4b show series of dI/dV spectra of the single-layer 1T'-WTe$_2$ with different coverage of surface K. The corresponding surface morphologies can be found in the supplementary materials Fig. S5. The features as marked by triangles in Figs. 4a,b can be used to determine the energy shift of the Fermi level, and their dependence on the K coverage is plotted in Fig. 4d. The maximal upward shift of the Fermi level is ~150 mV. Surprisingly, as shown in Fig. 4b,c, the energy gap is always pinned at the E$_F$ regardless of the position of E$_F$, indicating that the gap may arise from the electron-electron Coulomb interaction. Such a gap can persist at elevated temperatures up to ~75 K, as demonstrated in our tunneling spectroscopy measurement of supplementary materials Fig. S6.

As early as the 1970s, it was proposed that in localized systems the long-range Coulomb interactions between electrons diminish the single-particle DOS in the vicinity of the Fermi level, and thus a soft gap in the DOS is formed, which was called the Coulomb gap [26-30]. In the 2D case, the DOS near the Fermi energy can be qualitatively given as [30,31]:

$$G(\varepsilon) \propto (2/\pi e^4)|\varepsilon|,$$

at T = 0 K, where ε represents the energy with respect to the Fermi energy $E_F$. The Coulomb gap in the DOS can be observed experimentally at low enough temperatures, such that thermal excitations do not wash it out. Due to its sufficient decoupling from the BLG/SiC substrate [20], the single-layer WTe$_2$ is expected to form an ideal 2D localized electron system. The distorted W atomic zig-zag chains further enhance the anisotropy of localization. For our tunneling spectroscopy measurement, the dI/dV curves near the E$_F$ can be fitted very well with the linear format of a soft gap, as shown in Fig. 4c [30,31]. Considering the case that the gap is in a linear shape and always pinned at the E$_F$, a Coulomb gap is strongly suggested.

Basically, the Coulomb gap can be understood as a result of exitonic attraction of electrons and holes near the Fermi level, which depletes the one-particle DOS there [31]. Consider a transfer of one electron from a site *i* that is occupied in the ground state to a site *j* that is vacant in the ground state. The resulting energy change is given by

$\Delta E = E_j - E_i - e^2/r_{ij}$, where $E_j$ ( $E_i$ ) is the single-particle energy at site *i* (*j*) and the last term describes "the exitonic effect", i.e., the Coulomb interaction of the created electron-hole pair [32] with r$_{ij}$ the distance between the sites *i* and *j*. According to the stability criteria of the ground state, $\Delta E$ must be positive. It can be shown [35] that the concentration of such sites, n(ε) , cannot exceed (ε/e$^2$)$^2$ in the 2D case. Thus, the 2D single-particle DOS, G(ε) = dn(ε)/dε, is proportional to ε, vanishing when ε tends to zero at least as fast as ε.

As determined by the QPI characterization, the Fermi level of the undoped single-layer 1T'-WTe$_2$ crosses the two conduction bands along the Y-Γ-Y direction, and cuts the valence band in proximity of its top. In the K doping regime, the Fermi level is shifted upwards, above the top of the valence band, and thus the Fermi surface is composed of the two electron-like pockets. Following the theoretical model [31], our experimental results indicate that in all the cases, the Coulomb gap appears at $E_F$, in the conduction bands and/or in the valence band. The opening of Coulomb gap results in the insulating behavior of 1T'-WTe$_2$ at low temperature.

In summary, we have characterized the single-layer 1*T*'-WTe$_2$ with high resolution QPI-STS/STM, and verified a semimetal band structure, where there exists a band inversion near the Γ point, without a full SOC-induced bulk band gap. An important finding is the Coulomb gap induced by the electron interactions in this 2D localized electron system. To the best of our knowledge, it is the first time to experimentally observe the Coulomb gap in the single-layer QSH systems, which is vital to distinguish the topological edge states from the vanishing bulk band state and greatly facilitates the realization of QSH effect. Further in-depth experimental studies are strongly demanded.

Note added, during the manuscript preparation, we became aware a relevant work which realized the QSH effect in single-layer 1T'-WTe$_2$ at high temperature up to 100 K [32].

**ACKNOWLEDGMENTS**

This work was supported by the Ministry of Science and Technology of China (Grants No. 2014CB921103, No. 2013CB922103), the National Natural Science Foundation of China (Grants No. 11774149, No. 11374140, No. 11674165), the Fundamental Research Funds for the Central Universities (No. 020414380038), the National Thousand-Young-Talents Program (No. 0204128006), and the Open Research Fund Program of the State Key Laboratory of Low-Dimensional Quantum Physics.

**Methods**

**Sample preparation and STM/STS characterization.** The nearly freestanding single-layer 1T'-WTe$_2$ film was prepared by MBE technique on the bilayer graphene (BLG) formed on the 6*H*-SiC(0001) substrate. The detailed procedure of the sample growth can be found elsewhere [20]. After the MBE growth, the sample was transferred immediately into the LT-STM (Unisoku Co., USM1600) for scan at ~4.2 K. Differential conductance dI/dV spectra were acquired through a standard lock-in technique with the ac modulation of ~3-5 mV at 996 Hz. Experimental QPI maps were generated by symmetrizing the Fourier transformed dI/dV maps.

**FIGURE CAPTIONS**

**FIG. 1 STM topography and dI/dV spectra of single-layer 1T'-WTe$_2$ grown on BLG/SiC(0001)**. (a) Atomic model of the single-layer 1T'-WTe$_2$ structure grown on bilayer graphene /SiC substrate. (b) Atomic resolution STM image of the single-layer 1T'-WTe$_2$ surface (5 × 5 nm$^2$, U = +15 mV, I$_t$ = 300 pA). The length of the scale bar is 1 nm. The right half of the STM image is overlaid by the lattice of 1T'-WTe$_2$ with the top-layer Te atoms represented by yellow balls. (c) The corresponding Brillion Zone of single-layer 1T'-WTe$_2$. (d) Spatial variation of dI/dV spectra (U= +100 mV, I$_t$ = 200 pA) taken at 128 locations along a line of ~16 nm. (e,f) Two typical dI/dV maps (18 × 18 nm$^2$) taken on the single-layer 1T'-WTe$_2$ terrace, with bias of +70 mV (e) and -100 mV (f), respectively. The QPI-induced spatial modulations are clearly identified.

**FIG. 2 QPI patterns and the FFT images at different bias energies**. (a-d) Series of FFT images transformed from the measured dI/dV maps, with the bias voltages of +100 mV, +40 mV, -40 mV and -80 mV respectively. The dI/dV maps are measured on the same region as shown in Figs. 1e and 1f. The colored ovals represent the scattering $q$ vectors as indicated in (e). All the FFT images are symmetrized and drift-corrected. (e) Schematic band structure along the Y-Γ-Y direction. The intra- and inter- band scatterings are marked by the red arrows. The $q_1$ represents the intra-band scattering of the conduction band, which forms the pink ovals (a-d). The $q_2$ represents the inter-band scattering between the two conduction bands, which forms the green ovals in (a-d). The $q_3$ represents the inter-band scattering between the valence band and conduction band, which forms the orange oval in (c-d). The $q_4$ represents the intra-band scattering of the valence band, which forms the red oval in (d). (f) E-q dispersion along the Y-Γ-Y direction. In the region labeled by 2D-FFT, from -80 mV to +100 mV, the dispersion is extracted from the line-cut profiles in the 2D FFT images in the supplementary materials Fig. S2. The position of the line-cut is marked by the black line in (a). In the regions labeled by 1D-FFT, from -200 mV to

-80 mV and from +100 mV to +200 mV, the dispersion is extracted from the 1D-FFT of the dI/dV spectroscopic map taken along the *y* axis (Fig. S3). The black lines schematically illustrate the band dispersion of $q_2$, $q_3$ and $q_4$.

**FIG. 3 Comparison of the QPI patterns with the DFT simulation.** (a,b) The fast Fourier transform (FFT) images transformed from the dI/dV maps taken at +70 mV and -80 mV respectively. The FFT images are symmetrized and drift-corrected. Only the patterns in the first Brillion Zone are shown. The overlaid yellow dashed ovals briefly outline the QPI patterns. (c,d) Schematic constant energy contours at $E = +100$ mV and $E = -70$ mV, respectively. (e,f) The DFT simulated QPI patterns based on the CEC of (c,d).

**FIG. 4. dI/dV spectra taken on the 1$T$'-WTe$_2$ surface with different potassium (K) coverage.** (a), (b) dI/dV spectra taken in a large and small energy scales. The black colored triangles in (a) and white colored triangles in (b) mark the characteristic features in the dI/dV spectra, which move towards the left upon K deposition and are used to determine the shift of Fermi energy. The gap region is marked by the red vertical lines in (b). (c) The Zoom-in spectra of (b) showing the DOS near the Fermi energy. The suppression of the DOS at the Fermi level is always distinguishable. The background is subtracted, and the DOS is fitted with a linear function near the Fermi energy. (d) The position of the characteristic features as marked in (a,b) as a function of K coverage.

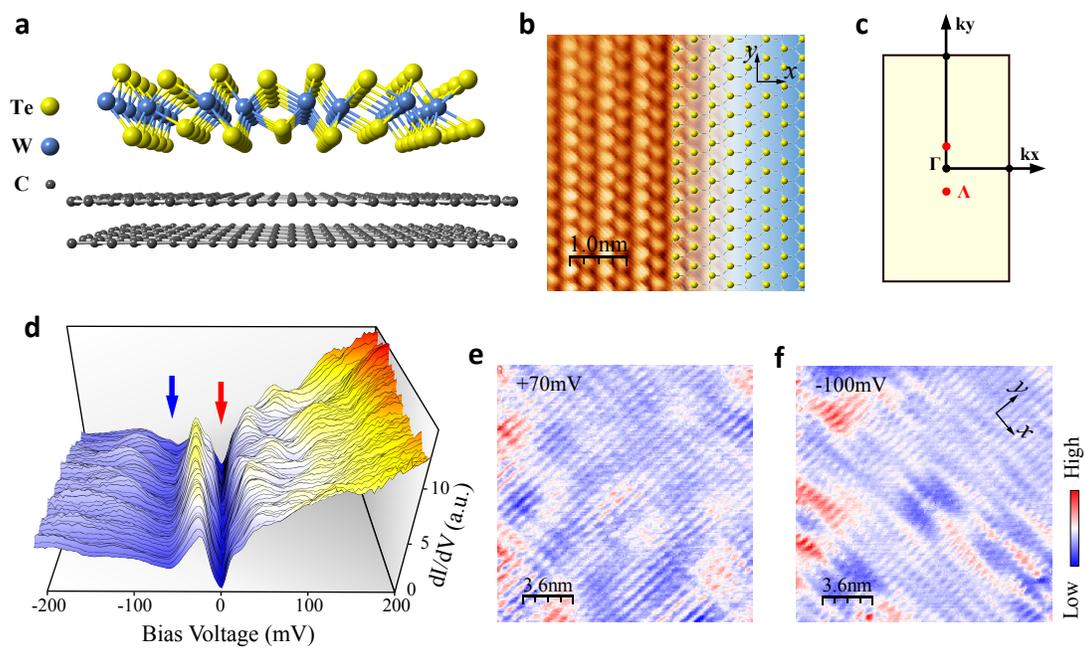

Figure 1

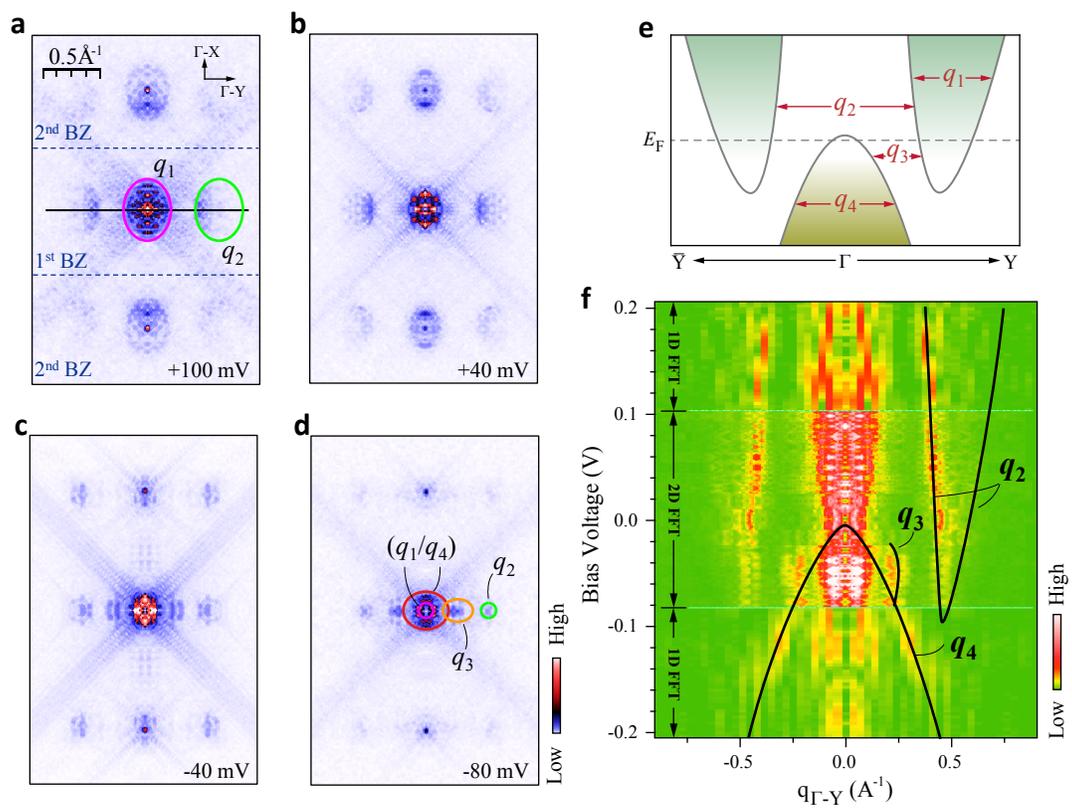

Figure 2

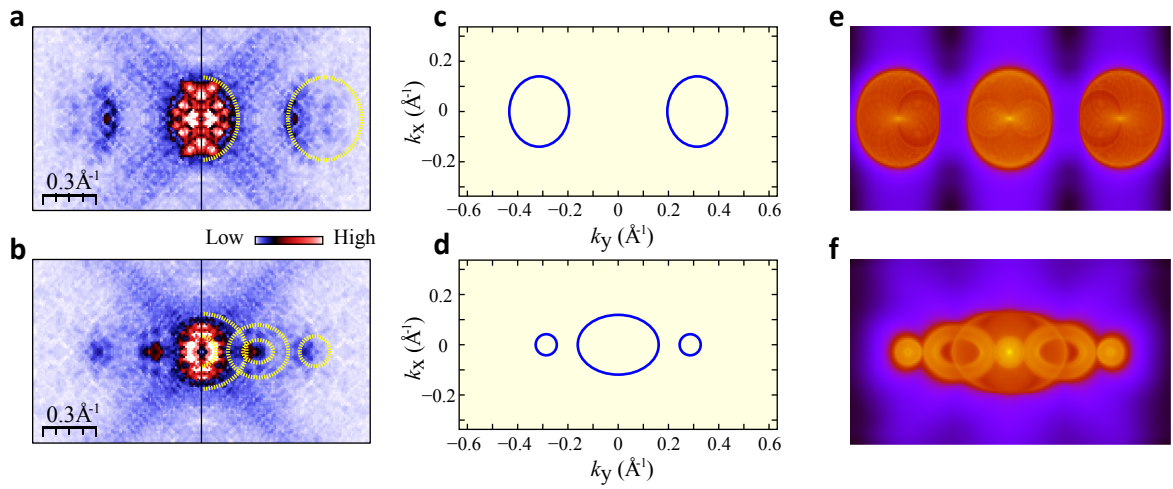

Figure 3

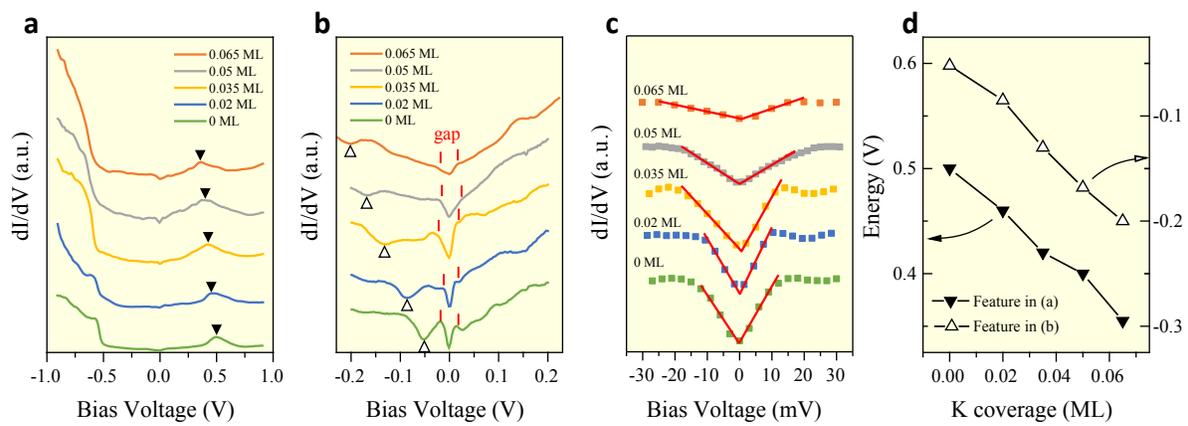

Figure 4